\documentclass[11pt]{amsart}
\usepackage{amssymb}
\usepackage{amsfonts}
\usepackage{float}
\usepackage{lmodern}
\usepackage{color}
\usepackage{graphicx}
\usepackage{lmodern}
\usepackage[paper=a4paper]{geometry}

\makeatletter

\@addtoreset{equation}{section}
\makeatother

\newcommand{\e}{\mathrm{e}}

\newcommand{\E}{\mathbb{E}}

\theoremstyle{plain}

\numberwithin{equation}{section}

\makeatletter
\g@addto@macro{\endabstract}{\@setabstract}
\newcommand{\authorfootnotes}{\renewcommand\thefootnote{\@fnsymbol\c@footnote}}
\makeatother

\begin{document}
\title{Wehrl entropies and Euclidean Landau levels}
\author{Z. MOUAYN$^{\ast}$, H. KASSOGUE$^{\diamond}$, P. KAYUPE KIKODIO$%
^{\sharp}$ AND I. F. FATANI$^{\flat}$}
\maketitle

\begin{abstract}
We are concerned with an information-theoretic measure of uncertainty for
quantum systems. Precisely, the Wehrl entropy of the phase-space probability 
$Q^{(m)}_{\hat{\rho}}=\left\langle z,m|\hat{\rho}|z,m\right\rangle $ which
is known as Husimi function, where $\hat{\rho}$ is a density operator and $%
\left|z,m\right\rangle $ are coherent states attached to an Euclidean $m$th
Landau level. We obtain the Husimi function $Q^{(m)}_{\beta}$ of the thermal
density operator $\hat{\rho}_{\beta}$ of the harmonic oscillator, which
leads by duality, to the Laguerre probability distribution of the mixed
light. We discuss some basic properties of $Q^{(m)}_{\beta}$ such as its
characteristic function and its limiting logarithmic moment generating
function from which we derive the rate function of the sequence of
probability distributions $Q^{(m)}_{\beta},\ m=0,1,2,...$. For $m\geq1$, we
establish an exact expression for the Wehrl entropy of the density operator $%
\hat{\rho}_{\beta}$ and we discuss the behavior of this entropy with respect
to the temperature parameter $T=1/\beta$.
\end{abstract}

%

\section{Introduction}

A coherent state $\left\vert \alpha \right\rangle $, $\alpha \in X$, is a
specific overcomplete family of normalized vectors in {a} Hilbert space $%
\mathcal{H}$ corresponding to {quantum phenomena and provide} $\mathcal{H}$
with the following resolution of identity

\begin{equation}
1_{\mathcal{H}}=\int\limits_{X}\left\vert \alpha \right\rangle \left\langle
\alpha \right\vert d\mu \left( \alpha \right) \text{,}  \label{1.1}
\end{equation}%
where $X$ denotes the phase-space domain and $d\mu \left( \alpha \right) $
is an associated integration measure on $X$. They have long been known for
the harmonic oscillator, whose properties have been used as a model for more
general constructions \cite{Gazeau}. In such systems of {states,} the
density operator $\hat{\rho}$ for an arbitrary (pure or mixed) state can be
represented by the Husimi's $Q$-function \cite{Husimi}:

\begin{equation}
Q_{\hat{\rho}}(\alpha )=\left\langle \alpha \right\vert \hat{\rho}\left\vert
\alpha \right\rangle.  \label{1.2}
\end{equation}

It follows that $Q_{\hat{\rho}}(\alpha )$ is normalized, since

\begin{equation}
1=Tr\left[\hat{\rho}\right]=\int_{X}Q_{\hat{\rho}}(\alpha )d\mu (\alpha ). 
\tag{1.3}
\end{equation}

As the $Q$-function is introduced in a way that guarantees it to be non
negative, then it allows the representation of quantum states by a \textit{%
probability distribution} in the phase-space $X$. It provides, equivalently
to the Glauber-Sudarshan or Wigner representations, a basis for a formal
equivalence between the quantum and classical descriptions of optical
coherence \cite{Glauber}. Combining the notion of coherent states and the $Q$
representation of the density operator $\hat{\rho}$, Wehrl \cite{Wehrl}
introduced a notion of entropy associated with a quantum state of the system
under consideration as%
\begin{equation}
S_{W}(\hat{\rho})=-\int_{X}Q_{\hat{\rho}}(\alpha )LogQ_{\hat{\rho}}(\alpha
)d\mu (\alpha )\text{.}  \tag{1.4}  \label{usimi}
\end{equation}%
This is the analogue of the Shannon entropy where the summation is replaced
by the integration over the phase-space $X$. Wehrl entropy cannot be
negative because $0\leq Q_{\hat{\rho}}(\alpha )\leq 1$ and the condition $%
\left( 1.3\right) $. Moreover, Wehrl proved that for, a given state $\hat{%
\rho}$, its entropy is greater than the von {Neumann} entropy $S_{N}(\hat{%
\rho})=-Tr\left( \hat{\rho}\log \hat{\rho}\right) $, which is zero for all
pure states regardless of whether they are coherent or not. In contrast,
Wehrl entropy clearly distinguishes coherent states and can be used as a
measure of the {strength} of the coherent component\ in a given quantum
state \cite{Orlowski}. The important inequality $S_{W}(\hat{\rho})\geq 1$, {%
first conjectured by Wehrl \cite{Wehrl} and then proved by Lieb \cite{Lieb}} 
{is a manifestation of the uncertainty principle.} The equality $S_{W}(\hat{%
\rho})=1$ holds if and only if $\hat{\rho}$ is the density operator of a
coherent state.

In this paper, we are concerned with the Husimi function $Q_{\hat{\rho}%
}^{(m)}:=\left\langle z,m|\hat{\rho}|z,m\right\rangle $ where $z\in \mathbb{C%
}$, $\hat{\rho}$ is a density operator and $\left\vert z,m\right\rangle $ is
a set of coherent states attached to higher $m$th Euclidean Landau level 
\cite{Mouayn,Mouayn2,Mouayn3}. {These coherent states can be obtained via a
group theoretical method \cite{Perelomov} or as series expansions where
coefficients ${z^{n}}/\sqrt{n!}$ of the canonical coherent states are
replaced by polyanalytic coefficients \cite{Mouayn2}, \cite{Abreusampling}.
This construction leads to the displaced Fock states \cite{Kral}, which are
widely used in quantum optics (see \cite{Dodonov} and references therein)}.
With respect to this set of coherent states, we first obtain the $Q$%
-function for a pure state. This enables us to get the $Q$-representation $%
Q_{\beta }^{(m)}$ of the thermal density operator $\hat{\rho}_{\beta }$ for
the harmonic oscillator, which leads, by duality, to the well-known Laguerre
distribution of the mixed light. We also discuss some basic properties of $%
Q_{\beta }^{(m)}$, namely: the characteristic function, the limiting
logarithmic moment generating function and the rate function of the sequence
of distributions $Q_{\beta }^{(m)},\ m=0,1,2,...$. For each fixed $m$, we
establish an exact expression for the Wehrl entropy of the density operator $%
\hat{\rho}_{\beta }$ and we discuss the behavior of this entropy with
respect to the temperature parameter $T=1/\beta $. For $m\geq 1$, we find
the precise temperature for which the Wehrl entropy attains {its} minimum.

The paper is organized as follows. In Section 2, we recall the construction
of coherent states attached to Euclidean Landau levels. In Section 3, we
give the Husimi function for a pure state. This expression is then used to
calculate the Husimi function for the thermal density operator $\hat{\rho}%
_{\beta}$ for the harmonic oscillator. In Section 4, we establish an exact
expression for the Wehrl entropy of $\hat{\rho}_{\beta}$ and we discuss its
asymptotic behavior with respect to the temperature parameter $T=1/\beta$.

\section{Coherent states attached to Landau levels}

The concept of generalized coherent states is used here following
Perelomov's presentation \cite{Perelomov}. The simplest operators used {to}
describe a quantum mechanical system with one degree of freedom are the
coordinate operator $q$ and the momentum operator $p$. Together with the
identity operator $I$, they satisfy the commutation relations: $%
[p,q]=-i\hbar I,\ \ [p,I]=[q,I]=0$, which characterize the well-known
Heisenberg-Weyl Lie algebra. In this algebra, an element $W$ of the form $%
W=u(ip)+v(iq)+t(iI),\ u,\ v,\ t\in \mathbb{R}\ \ \ (\hbar =1)$ can be
rewritten in terms of annihilation and creation operators 
\begin{equation}
a=\frac{1}{\sqrt{2}}(q+ip),\ \ \ a^{\ast }=\frac{1}{\sqrt{2}}(q-ip)
\label{a_operators}
\end{equation}%
%
%
%
as follows $W=t(iI)+\alpha a^{\ast }-\alpha ^{\ast }a$, where 
\begin{equation}
\alpha =\frac{1}{\sqrt{2}}(-u+iv),\ \ \ \alpha ^{\ast }=\frac{1}{\sqrt{2}}%
(-u-iv).
\end{equation}%
%
By exponentiation, we obtain that $\exp (W)=\exp (itI)D(\alpha )$ and $%
D(\alpha )=\exp (\alpha a^{\ast }-\alpha ^{\ast }a).$ The multiplication law
for operators $D(\alpha )$ has the form $D(\alpha )D(\beta )=\exp (i\Im
\alpha \beta ^{\ast })D(\alpha +\beta )$. As a consequence, the operators $%
T(t,\alpha ):=\exp (it)D(\alpha )$ form an unitary irreducible
representation (UIR) of the Heisenberg-Weyl Lie group $\mathbb{W}_{1}$ whose
underlying manifold is $\mathbb{R}\times \mathbb{C}$. Applying $%
T(g)=T(0,\alpha )$ to a vector $\left\vert \phi _{0}\right\rangle $ of $%
L^{2}(\mathbb{R})$, we obtain a set of states $\left\vert \alpha
\right\rangle :=D(\alpha )\left\vert \phi _{0}\right\rangle $. The set ${%
\left\vert \alpha \right\rangle }$ is a system of generalized coherent
states (GCS), which satisfy (\cite{Perelomov}, p.15): 
\begin{equation}
\left\langle \psi ,\psi \right\rangle =\int_{\mathbb{C}}d\mu (\alpha
)\left\langle \psi|\alpha \right\rangle \left\langle \alpha |\psi
\right\rangle ,\ \ \ \psi \in L^{2}(\mathbb{R}),  \label{spi_identity}
\end{equation}%
%
%
%
$d\mu\left(\alpha\right)\propto d\alpha_{x}d\alpha_{y}$ is proportionnal to
the usual planar measure, where $\alpha=\alpha_x+i\alpha_y$. 
We shall use the real variables $x:=B^{1/2}u$ and $y:=B^{1/2}v$. In terms of 
$(x,y)$, the action of operator $D(\alpha ):=D(x,y)$ on functions $\psi \in
L^{2}(\mathbb{R})$ takes the following form 
\begin{equation}
D(x,y)[\psi ](\xi )=\exp \left( i\frac{B}{2}xy\right) \exp \left( -i\sqrt{B}%
y\xi \right) \psi (\xi -\sqrt{B}x),\ \ \xi \in \mathbb{R}.  \label{D_psi}
\end{equation}%
%
%
%
Now, we choose as vector $\left\vert \phi _{0}\right\rangle $ the function
defined by 
\begin{equation}
\Phi _{m}(\xi )=(\sqrt{\pi }2^{m}m!)^{-\frac{1}{2}}e^{-\frac{1}{2}\xi
^{2}}H_{m}(\xi ),  \label{Phi_m}
\end{equation}%
%
%
%
where $H_{m}$ is the $m$th Hermite polynomial \cite{Gradshteyn}, to define
coherent states labeled by elements $(x,y)\in \mathbb{R}^{2}$ as $\left\vert
(x,y),B,m\right\rangle :=D(x,y)[\Phi _{m}].$ By \eqref{Phi_m} and %
\eqref{D_psi}, the wave functions of these GCS have the form 
\begin{equation}
\left\langle \xi |(x,y),B,m\right\rangle =(\sqrt{\pi }2^{m}m!)^{-\frac{1}{2}%
}\exp \left( -i\sqrt{B}\xi y+i\frac{B}{2}xy-\frac{1}{2}(\xi -\sqrt{B}%
x)^{2}\right) H_{m}(\xi -\sqrt{B}x)\text{.}  \label{cscs}
\end{equation}%
%
%
%
For these GCS, equation \eqref{spi_identity} reads 
\begin{equation}
\int_{\mathbb{R}^{2}}d\mu (x,y)\left\langle \psi |(x,y),B,m\right\rangle
\left\langle (x,y),B,m|\psi\right\rangle =\left\langle \psi ,\psi
\right\rangle ,\ \ \psi \in L^{2}(\mathbb{R})
\end{equation}%
%
%
%
with $d\mu (x,y)=\frac{B}{2\pi }dxdy$. This implies the resolution of the
identity of the Hilbert space $L^{2}\left( {\mathbb{R}}\right) $ as 
\begin{equation}
\mathbf{1}_{L^{2}\left( {\mathbb{R}}\right) }=\int_{{\mathbb{R}}^{2}}\mid
(x,y),B,m\rangle \langle (x,y),B,m\mid d\mu (x,y).
\end{equation}%
%
%
%
With the coherent states \eqref{cscs} one can associate the coherent state
transform $\mathcal{W}_{B,m}:L^{2}({\mathbb{R}},d\xi )\rightarrow L^{2}({%
\mathbb{R}^{2}},d\mu )$ defined as 
\begin{equation}
\mathcal{W}_{B,m}[\psi ](x,y):=\int_{\mathbb{R}}d\xi \psi (\xi )\overline{%
\left\langle \xi |(x,y),B,m\right\rangle} .
\end{equation}%
%
%
%
Thanks to the square integrability of the UIR $D(x,y)$, this transform is an
isometrical map whose range \cite{Mouayn3} is 
\begin{equation}
\mathcal{E}_{B,m}:=\mathcal{W}_{B,m}\left[ L^{2}({\mathbb{R}})\right]
=\left\{ \varphi \in L^{2}\left( \mathbb{R}^{2},dxdy\right) ,H_{B}\left[
\varphi \right] =\epsilon _{m}\varphi \right\}  \label{espace}
\end{equation}%
%
%
%
where 
\begin{equation}
H_{B}=\frac{1}{2}\left( \left( i\frac{\partial }{\partial x}-\frac{B}{2}%
y\right) ^{2}+\left( i\frac{\partial }{\partial y}+\frac{B}{2}x\right)
^{2}\right)
\end{equation}%
%
%
%
is the Landau Hamiltonian acting in the Hilbert space $L^{2}\left( \mathbb{R}%
^{2},d\mu \right) $, whose spectrum of $H_{B}$ consists of infinite number
of eigenvalues with infinite multiplicity of the form 
\begin{equation}
\epsilon _{m}=\left( m+\frac{1}{2}\right) B,\quad m=0,1,2,...\ ,
\end{equation}%
%
%
%
the well known Euclidean Landau levels.\newline
\newline
\textbf{Remark 2.1.} In \cite{Abreu}, completeness properties of discrete
coherent {states} $\left\vert z,m\right\rangle $ associated with higher
Euclidean Landau levels have been investigated and classical results of
Perelomov and of Bargmann, Butera, Girardello and Klauder were extended
using the structure of Gabor spaces \cite{AbreuGabor}, which allows to
identify the completeness problem with known results from {Gabor theory}.
Using the recent characterization of complete Gabor systems obtained in \cite%
{Gabor}, a full description of the lattices with rational density.

\section{The $Q$ representation in coherent states $\left\vert(x,y),B,m%
\right\rangle $}

\subsection*{Husimi's Q-function for a pure state}

Let $B>0$ and $m,j\in \mathbb{Z}_{+}$ be fixed parameters. The Husimi
function of the pure state projector $\hat{\rho}_{j}=|j\rangle \langle j|$
on a Fock state $\left|j\right\rangle$ is defined by 
\begin{equation}
Q_{\hat{\rho}_{j}}((x,y),B,m)=|\langle (x,y),B,m|j\rangle |^{2},\ \ \
(x,y)\in \mathbb{R}^{2}\text{.}
\end{equation}%
Explicitly, 
\begin{equation}
Q_{\hat{\rho}_{j}}((x,y),B,m)=\frac{(m\wedge j)!}{(m\vee j)!}e^{-\frac{B}{2}%
(x^{2}+y^{2})}\left( \frac{B}{2}(x^{2}+y^{2})\right) ^{|m-j|}\left(
L_{m\wedge j}^{(|m-j|)}\left( \frac{B}{2}(x^{2}+y^{2})\right) \right) ^{2}%
\text{,}  \label{Husimi}
\end{equation}%
%
%
%
where quantities $m\wedge j$ and $m\vee j$ denote respectively the minimal
and the maximal value between $m,j$ and $L_{n}^{(\alpha )}(.)$ denotes the $n
$th Laguerre polynomial \cite{Gradshteyn}. See Appendix \ref{appA} for the
proof of \eqref{Husimi}.

To simplify the notation, we set $Q_{j}^{(m)}\equiv Q_{\hat{\rho}%
_{j}}((x,y),B,m)$. Several properties of this probability distribution can
be derived from its characteristic function (the proof is given in Appendix %
\ref{appB}). \newline
\ \newline
\textbf{Proposition 3.1. }\textit{The characteristic function of }$%
Q_{j}^{(m)}$\textit{\ has the following expression} 
\begin{equation}
\mathbb{E}\left( e^{iuQ_{j}^{(m)}}\right) =\frac{(m+j)!}{m!j!}%
(1-iu)^{-(m+j+1)}\ _{2}F_{1}\left( 
\begin{array}{c}
-m,-j \\ 
-m-j%
\end{array}%
\mid 1+u^{2}\right) ,\ \ u\in \mathbb{R}.
\end{equation}%
%
%
%
\textit{In particular, the mean value and the variance are respectively} $%
\mathbb{E}(Q_{j}^{(m)})=m+j+1$ \textit{and }$\mathbb{V}%
ar(Q_{j}^{(m)})=2mj+m+j+1$. \newline
\ \newline
\textbf{Remark 3.1.} It is clear that $Q_{j}^{(m)}$ vanishes at the origin $%
(0,0)$ and at points $(x,y)\in \mathbb{R}^{2}$ such that $B(x^{2}+y^{2})/2$
is a zero of the Laguerre polynomial $L_{m\wedge j}^{(|m-j|)}(.)$. We denote
by $x_{i}^{(m\wedge j)},1\leq i\leq m\wedge j$, the ordered consecutive
positive zeros of this polynomial. Then the zeros of $Q_{j}^{(m)}$ are
located on concentric circles of radius $R_{i}=\sqrt{\frac{2}{B}%
x_{i}^{(m\wedge j)}}$. When $m=0$, we set $\lambda =(x^{2}+y^{2})B/2$ and
the expression $(\ref{Husimi})$ becomes 
\begin{equation}
Q_{j}^{(0)}(\lambda )=\frac{1}{j!}e^{-\lambda }\lambda ^{j}\text{,}
\end{equation}%
the Gamma probability density with mean $(j+1)$ for each fixed $j=0,1,...$.
A discussion on the number of zeros of the Husimi density and their location
in connection with properties of harmonic and anaharmonic oscillators can be
found in \cite{Korsh}. \newline
\newline
\textbf{Remark 3.2.} For $m\geq 1$, we note that 
\begin{equation}
Q_{j}^{(m)}(\lambda )=\frac{(m\wedge j)!}{(m\vee j)!}e^{-\lambda }\lambda
^{|m-j|}\left( L_{m\wedge j}^{(|m-j|)}(\lambda )\right) ^{2}  \label{rhomj}
\end{equation}%
%
is a Rakhmanov probability density of Laguerre-type \cite{Dehesa}. We also
observe that $Q_{j}^{(m)}$ coincides with the random variable, denoted $%
S_{j}^{(m,1)}$, and defined by Shirai \cite{Shirai} in the context of
determinantal processes for higher Landau levels (see \cite{APRT} for an
alternative approach and \cite{HendHaimi,HaiHen2} for finite-dimensional
versions) : 
\begin{equation}
\text{P}r\left( S_{j}^{(m,1)}\leq r^{2}\right) =\kappa _{j}^{(m,1)}(r^{2})
\end{equation}%
%
as a probabilistic representation of eigenvalues $\kappa _{j}^{(m,1)}(r^{2})$
of the restricted operator $\left( K_{m,B}\right) _{D_{r}}$ to the disk $%
D_{r}\subset \mathbb{C}$ of radius $r$, where $K_{m,B}:L^{2}\left( \mathbb{R}%
^{2}\right) \rightarrow \mathcal{E}_{B,m}$ is the projection operator onto
the $m$th Landau eigenspace $\mathcal{E}_{B,m}$ in $(\ref{espace})$. From
this observation and the results in \cite{Shirai}, it follows that the
limiting logarithmic moment generating function of the $\left(
Q_{j}^{(m)}\right) _{j\geq 0}$ distributions is 
\begin{equation}
\wedge _{j}(u):=\lim_{j\rightarrow +\infty }\frac{1}{j}\log \mathbb{E}\left(
e^{uQ_{j}^{(m)}}\right) =-\log (1-u),\ u<1.
\end{equation}%
From the latter one, the rate function $\wedge _{j}^{\ast }(\xi )=\xi
-1-\log \xi $, was deduced via a Legendre transformation. \newline
\newline
\textbf{Remark 3.3.} By duality, to $Q_{j}^{(m)}$ a non-negative
integer-valued random variable $X_{\lambda }^{(m)}$ are associated as 
\begin{equation}
\text{P}r\left( X_{\lambda }^{(m)}=j\right) =Q_{j}^{(m)}(\lambda ),\ \
j=0,1,2,...\ .
\end{equation}%
For $m=0$, $X_{\lambda }^{(0)}$ is the Poisson distribution and for $m\geq 1$%
, nonclassical properties of the coherent states $\left\vert
z,m\right\rangle $ were investigated in \cite{Mouayn4} through statistics of 
$X_{\lambda }^{(m)}$ as a photon count probability distribution. The
analysis of $X_{\lambda }^{(m)}$ can be found in \cite{Mouayn5} where it was
proved that for $m\geq 1$ it is not infinitely divisible in contrast with $%
X_{\lambda }^{(0)}$ and a L\'{e}vy-Khintchine-type representation of its
characteristic function was also derived.

\subsection{Husimi's Q-function for a thermal density operator.}

The standard statistical mechanics starts using the Gibb's canonical
distribution, whose thermal density operator is represented by 
\begin{equation}
\hat{\rho}_{\beta }=\frac{1}{Z}e^{-\beta \hat{H}}
\end{equation}%
where $Z=Tr\left( e^{-\beta \hat{H}}\right) $ is the partition function, $%
\hat{H}$ is the Hamiltonian of the system, $\beta =1/(kT)$ the inverse
temperature $T$ and $k$ the Boltzmann constant ($k=1$). In our context, 
\begin{equation}
\hat{H}:=-\frac{d^{2}}{d\xi^{2}}+\xi^{2}
\end{equation}%
is the harmonic oscillator and $\hat{\rho}_{\beta }$ is the normalized heat
operator associated with $\hat{H}$ 
\begin{equation}
\hat{\rho}_{\beta }=\frac{1}{Z}\sum_{j=0}^{+\infty }e^{-\beta \left( j+\frac{%
1}{2}\right) }\left\vert j\right\rangle \left\langle j\right\vert
\label{3.12}
\end{equation}%
where $\left\vert j\right\rangle $ are the eigenstates of $\hat{H}$. The
Husimi distribution is defined as the expectation value of the density
operator $\hat{\rho}_{\beta }$ in the set of coherent states $\left\vert
(x,y),B,m\right\rangle $ by 
\begin{equation}
Q^{(m)}_{\beta}((x,y),B):=\left\langle (x,y),B,m\right\vert \hat{\rho}%
_{\beta }\left\vert (x,y),B,m\right\rangle ,\text{ }(x,y)\in \mathbb{R}^{2}%
\text{.}  \label{3.122}
\end{equation}%
By using $(\ref{3.12})$, Eq.$(\ref{3.122})$ reads 
\begin{equation}  \label{Qb_m}
Q^{(m)}_{\beta}((x,y),B)=\frac{1}{Z}\sum_{j=0}^{+\infty }e^{-\beta \left( j+%
\frac{1}{2}\right) }Q^{(m)}_j((x,y),B)
\end{equation}%
where $Q^{(m)}_{j}$ is the Husimi function given in $(\ref{Husimi})$. 
\newline
\ \newline
\textbf{Proposition 3.2.} \emph{Let $B>0$ and $m\in\mathbb{Z}_{+}$ be fixed
parameters. Then, the Husimi function of the density operator $\hat{\rho}%
_{\beta }$ in the set of coherent states $\left\vert (x,y),B,m\right\rangle $
is given by} 
\begin{equation}
Q^{(m)}_{\beta}((x,y),B)=\left( 1-e^{-\beta }\right) \exp \left( -\frac{B}{2}%
(x^{2}+y^{2})(1-e^{-\beta })\right) e^{-m\beta }L_{m}^{(0)}\left(
-2B(x^{2}+y^{2})sh^{2}\frac{\beta }{2}\right)  \label{husssimi}
\end{equation}%
\emph{for every} $(x,y)\in\mathbb{R}^2$. \newline
\newline
See Appendix \ref{appC} for the proof. \newline
\newline
\textbf{Proposition 3.3.} \textit{The characteristic function of }$%
Q^{(m)}_{\beta}$\textit{\ has the following expression} 
\begin{equation}  \label{charactQbeta}
\mathbb{E}\left(e^{iu Q^{(m)}_{\beta}}\right)=\left(1-e^{-\beta}\right)\frac{%
\left(1-e^{-\beta}-iue^{-\beta}\right)^m}{\left(1-e^{-\beta}-iu\right)^{m+1}}%
,\ \ u\in \mathbb{R}.
\end{equation}%
\textit{In particular, the mean value and the variance are respectively} $%
\mathbb{E}(Q^{(m)}_{\beta})=m+\left(1-e^{-\beta}\right)^{-1}$ \textit{and }$%
\mathbb{V}ar(Q^{(m)}_{\beta})=\left(m+1-e^{-2\beta}\right)\left(1-e^{-\beta}%
\right)^{-2}$. \newline
\newline
\textbf{Proof.} Setting $\lambda=-B(x^2+y^2)/2$, the characteristic function
in $(\ref{husssimi})$ reads 
\begin{equation}
\mathbb{E}\left(e^{iuQ^{(m)}_{\beta}}\right)=\frac{\left(1-e^{-\beta}\right)%
}{e^{m\beta}} \int_{0}^{+\infty}exp\left((1-e^{-\beta}-iu)\lambda%
\right)L_m^{(0)}\left(-4\lambda\sinh^2\left(\frac{\beta}{2}\right)\right)
d\lambda.
\end{equation}
By using the formula (\cite{Brychkov}, p.151 ): 
\begin{equation}  \label{3.17}
\int_0^{+\infty}e^{-by}L_{\nu}^{(0)}(ay)dy=\frac{(b-a)^{\nu}}{b^{v+1}},\ \
\Re(b)>-1,\ \Re(b-a)>0,
\end{equation}
for parameters $b=1-e^{-\beta}-iu$, $a=-4\sinh^2(\beta/2)$ and $\nu=m$, we
arrive at $(\ref{charactQbeta})$.$\Box$\newline
\newline
\textbf{Corollary 3.1.} \emph{For each fixed $\beta>0$, the limiting
Logarithmic moment generating function of the Husimi density $%
Q^{(m)}_{\beta},\ m=0,1,2,...$, has the following expression } 
\begin{equation}
\wedge_{\beta}(u):=\lim_{m\to+\infty}\frac{1}{m}\log\mathbb{E}%
\left(e^{uQ_{\beta}^{(m)}}\right)=\log\left(\frac{1-e^{-\beta}-ue^{-\beta}}{%
1-e^{-\beta}-u}\right), \ u<1-e^{-\beta}.
\end{equation}
\ \newline
By direct calculations, we obtain the Fenchel-Legendre transform (\cite%
{Handa}, p.26) of the limiting logarithmic generating function $%
\wedge_{\beta}(u)$, which is defined by $\wedge^{\ast}_{\beta}(\xi)=%
\sup_{u<1-e^{-\beta}}\left(\xi u-\wedge_{\beta}(u)\right).$ \newline
\newline
\textbf{Proposition 3.4.} \emph{The rate function for the Husimi density } $%
Q^{(m)}_{\beta},\ m=0,1,2,...$, \emph{is given by } 
\begin{equation}
\wedge_{\beta}^{\ast}(\xi)=\xi u_{\xi}-\log\left(\frac{1-e^{-\beta}-u_{%
\xi}e^{-\beta}}{1-e^{-\beta}-u_{\xi}}\right)
\end{equation}
\emph{where} 
\begin{equation}
u_{\xi}= \frac{1}{2\xi e^{-\beta}}\left(\xi(1-e^{-\beta})-\sqrt{%
\xi^2(1-4e^{-\beta}+6e^{-2\beta}-4e^{-3\beta}+e^{-4\beta})+\xi(4e^{-%
\beta}-8e^{-2\beta}+4e^{-3\beta})}\right)
\end{equation}
\emph{for every} $\xi>0$. \ \newline
\newline
Finally, if we set $\lambda =(x^{2}+y^{2})B/2$ for fixed $x$ and $y$, then
by duality a non-negative integer-valued random variable $Y_{\lambda,\beta}$
is associated with $Q^{(m)}_{\beta}$ by 
\begin{equation}
\text{P}r\left(Y_{\lambda,\beta}=m\right)=Q^{(m)}_{\beta}(\lambda),\ \
m=0,1,2,...\ .
\end{equation}
In fact $Y_{\lambda,\beta}$ turns out to be the well-known Laguerre
probability distribution with parameter $(\lambda ,\beta )$. To present it
in a recognized form in the photon count theory, we recall that $T=1/k\beta $
represents, in suitable units $(k=1)$, the absolute temperature and $%
N_{T}:=\left( e^{\frac{1}{T}}-1\right) ^{-1}$ the average number of photons
associated with the thermal noisy state. With these notations, the photon
count probability distribution for a mixed light is given by [20-22]: 
\begin{equation}
\Pr (Y_{\lambda,\beta}=m)=\frac{(N_{T})^{m}}{(1+N_{T})^{m+1}}\exp \left( -%
\frac{{\lambda }}{1+N_{T}}\right) L_{m}^{(0)}\left( -\frac{\lambda }{%
N_{T}(1+N_{T})}\right) .  \label{3.16}
\end{equation}%
An interesting aspect of $(\ref{3.16})$ lies in the fact that when the
strength of the coherent light goes to zero $(\lambda \rightarrow 0)$ the
distribution reduces to the Bose-Einstein distribution while when the
thermal light decreases $(T\rightarrow 0)$ it converges to the Poisson
distribution of coherent a light.

\section{The Wehrl entropy for the harmonic oscillator in a thermal state}

In this section, we first proceed to determine the Wehrl entropy for a pure
state (see Appendix \ref{appD} for the proof). \newline
\newline
\textbf{Proposition 4.1.} \emph{The Wehrl entropy for the pure state} $\hat{%
\rho}_j=\left|j\right\rangle\left\langle j\right|$ \emph{is given by} 
\begin{equation}  \label{entrropy}
S_{W}(\hat{\rho}_{j})=\log\left( \frac{2\pi }{e}\frac{(m\wedge j)^{|m-j|+1}}{%
(m\vee j+1)^{|m-j|}}\right) +o(1)
\end{equation}%
\emph{and satisfies} 
\begin{equation}
S_{W}(\hat{\rho}_{j})\leq \log(\frac{2\pi }{e}(m\wedge j)),\ m,j=0,1,2,\dots
\ .
\end{equation}%
\newline
We observe that the entropy \eqref{entrropy} can also be obtained as a limit
as $q\rightarrow 1$ of the Renyi entropy of order $q$ of the Husimi density $%
Q^{(m)}_{j}$ in \eqref{Husimi} as 
\begin{equation}
S_{W}(\hat{\rho}_{j})=\lim_{q\rightarrow 1}R_{q}\left(
Q^{(m)}_{j}(\lambda)\right)
\end{equation}%
where 
\begin{equation}
R_{q}\left( Q^{(m)}_{j}(\lambda)\right) =\frac{1}{1-q}\log\left( \mathbb{E}%
\left( Q^{(m)}_{j}(\lambda)\right) ^{q-1}\right) .
\end{equation}%
Explicitly, for $2q\in \mathbb{N}$, with $q\neq 1$, we have 
\begin{equation*}
R_{q}(Q^{(m)}_{j}\left(\lambda\right)=\frac{1}{1-q}\log\Big[%
\sum_{k=0}^{2(m\wedge j)q}\frac{\Gamma (|m-j|q+k+1)}{q^{|m-j|q+k+1}}\frac{%
(2q)!}{(k+2q)!}
\end{equation*}%
\begin{equation}  \label{r_entropy}
\times B_{k+2q,2q}\left( c_{0}^{(m\wedge j,|m-j|)},2!c_{1}^{(m\wedge
j,|m-j|)},\dots ,(k+1)!c_{k}^{(m\wedge j,|m-j|)}\right) \Big]
\end{equation}%
where the coefficients $c_{k}^{(m\wedge j,|m-j|)}=0$ for $k>m\wedge j$ and 
\begin{equation}
c_{k}^{(m\wedge j,|m-j|)}=\sqrt{\frac{((m\wedge j)+k)!}{(m\wedge j)!}}\frac{%
(-1)^{k}}{(|m-j|+k)!}\binom{m\wedge j}{k}
\end{equation}%
otherwise. Here, 
\begin{equation}
B_{s,l}(a_{1},a_{2},\dots ,a_{s-l+1})=\sum \frac{s!}{j_{1}!j_{2}!\dots
j_{s-l+1}!}\left( \frac{a_{1}}{1!}\right) ^{j_{1}}\left( \frac{a_{2}}{2!}%
\right) ^{j_{2}}\dots \left( \frac{a_{s-l+1}}{(s-l+1)!}\right) ^{j_{s-l+1}}
\end{equation}%
where the sum runs over all of integer numbers $j_{1},j_{2},\dots ,j_{s-l+1}$
such that 
\begin{equation*}
j_{1}+j_{2}+\cdots +j_{s-l+1}=l,\quad j_{1}+2j_{2}+\cdots
+(s-l+1)j_{s-l+1}=s.
\end{equation*}%
are Bell polynomials (\cite{Comptet}, p.134]). The expression %
\eqref{r_entropy} can be obtained by simply adapting suitable parameters in
the formula (26) in ([14], p.1134) of the Renyi entropy. \newline
\newline
\textbf{Remark 4.1.} For $m=0$, the Husimi function in $(\ref{Husimi})$
reduces to $Q^{(0)}_{j}(\lambda)=(j!)^{-1}\lambda ^{j}e^{\lambda}$ where we
have set $\lambda=B(x^2+y^2)/2$. Direct calculations immediately gives (\cite%
{Orlowski}, p.287): 
\begin{equation}
S_{W}(\hat{\rho}_j)=1+j+\log(j!)-j\psi (j+1)
\end{equation}%
where $\psi (j+1)=-\gamma +\sum_{k=1}^{j}\frac{1}{k}$ and $\gamma \simeq
0.5772156649$ is the well-known Euler-Mascheroni constant.\newline
\newline
\textbf{Proposition 4.2.} \emph{The Wehrl entropy with respect to the set of
coherent states $\left\vert (x,y),B,m\right\rangle $ of the thermal density
operator $\hat{\rho}_{\beta }$ in $(\ref{3.12})$ has the following expression%
} 
\begin{equation}  \label{entttropy}
S_{W}(\hat{\rho}_{\beta})=1-\log(1-e^{-\beta })+m\left( \beta +e^{-\beta
}-e^{-2\beta }\right)
\end{equation}%
\emph{for} $m=0,1,2,...$ . \newline
\newline
\textbf{Proof.} The Wehrl entropy of the mixed state $\hat{\rho}_{\beta }$
is defined by 
\begin{equation}  \label{We}
\mathcal{S}_{W}(\hat{\rho}_{\beta}):=-\int_{\mathbb{R}^{2}}Q^{(m)}_{%
\beta}((x,y),B)\log\left[Q^{(m)}_{\beta}((x,y),B)\right]d\mu (x,y)
\end{equation}%
where%
\begin{equation*}
Q^{(m)}_{\beta}((x,y),B)=\eta e^{-m\beta }\exp \left( -\frac{B}{2}%
(x^{2}+y^{2}) \eta \right) L_{m}^{\left( 0\right) }\left( -\frac{B}{2}%
(x^{2}+y^{2})\eta^{2}e^{\beta }\right) ,
\end{equation*}%
with $\eta:=1-e^{-\beta }$. Using polar coordinates $x=r\cos \theta $, $%
y=r\sin \theta $, $r>0,\theta \in \lbrack 0,2\pi )$ and setting $u=\eta
^{2}e^{\beta }Br^{2}/2$, then we split the integral $(\ref{We})$ into a sum
of two integrals as%
\begin{equation*}
S_{W}\left(\hat{\rho}_{\beta}\right)=\frac{e^{-m\beta }}{(-\eta e^{\beta })}%
\log\left[ \epsilon e^{-m\beta }\right] \int_{0}^{+\infty }\exp \left( \frac{%
-u}{\eta e^{\beta }}\right) L_{m}^{(0)}(-u)du
\end{equation*}%
\begin{equation*}
+\frac{e^{-m\beta }}{(-\eta e^{\beta })}\int_{0}^{+\infty }\exp \left( \frac{%
-u}{\eta e^{\beta }}\right) L_{m}^{(0)}(-u)\log\left[ \exp \left( \frac{-u}{%
\eta e^{\beta }}\right) L_{m}^{(0)}(-u)\right] \,du.
\end{equation*}
By applying the formula $(\ref{3.17})$ with parameters $b=1/\eta e^{\beta
},\nu =m$ and $x=u$, we obtain 
\begin{equation}
\int_{0}^{+\infty }\exp \left( \frac{-u}{\eta e^{\beta }}\right)
L_{m}^{(0)}(-u)du=\eta e^{\beta }(1+\eta e^{\beta })^{m}.
\end{equation}%
So that the first integral reads 
\begin{equation}
I_{1}=-e^{-m\beta }\left( 1+\eta e^{\beta }\right) ^{m}\log\left( \eta
e^{-m\beta }\right) .  \label{I11}
\end{equation}%
The second integral 
\begin{equation}  \label{secondintegral}
I_{2}=\frac{e^{-m\beta }}{-\eta e^{\beta }}\int_{0}^{+\infty }\exp \left( 
\frac{-u}{\eta e^{\beta }}\right) L_{m}^{\left( 0\right) }(-u)\log\left[
\exp \left( \frac{-u}{\eta e^{\beta }}\right) L_{m}^{\left( 0\right) }(-u)%
\right]du
\end{equation}%
is written in the form 
\begin{equation}
I_{2}=\frac{e^{-m\beta }}{-\eta e^{\beta }}\frac{dG_{p}}{dp}\big{|}_{p=1},
\end{equation}%
where 
\begin{equation}
G_{p}:=\int_{0}^{+\infty }\exp \left( \frac{-pu}{\eta e^{\beta }}\right) %
\left[ L^{(0)}_{m}(-u)\right] ^{p}du.
\end{equation}%
The latter one may also be expressed as 
\begin{equation*}
G_{p}=\frac{\eta e^{\beta }}{p}\ F_{A}^{p}\left( 
\begin{array}{c}
1,-m,\cdots ,-m \\ 
1,\cdots ,1%
\end{array}%
\mid -\frac{\eta e^{\beta }}{p},\cdots ,-\frac{\eta e^{\beta }}{p}\right)
\end{equation*}%
in terms of the Lauricella hypergeometric function $F_{A}^{p}$ of $p$
variables by making use of the Mayr-Erd\'{e}lyi formula (\cite{Srivastava},
p.250): 
\begin{eqnarray}
\int_{0}^{\infty }x^{\rho -1}e^{-\sigma x}L_{m_{1}}^{(\alpha _{1})}(\lambda
_{1}x)\cdots L_{m_{r}}^{(\alpha _{r})}(\lambda _{r}x)dx &=&\binom{%
m_{1}+\alpha _{1}}{m_{1}}\cdots \binom{m_{r}+\alpha _{r}}{m_{r}}\frac{\Gamma
(\rho )}{\sigma ^{\rho }}  \notag \\
&&\times \ F_{A}^{r}\left( 
\begin{array}{c}
\rho ,-m_{1},\cdots ,-m_{r} \\ 
\alpha _{1}+1,\cdots ,\alpha _{r}+1%
\end{array}%
\mid \frac{\lambda _{1}}{\sigma },\cdots ,\frac{\lambda _{1}}{\sigma }%
\right) ,
\end{eqnarray}%
$\Re \sigma >0$ and $\Re \rho >0$, for parameters $p=1$,$\ \sigma =p/\eta
e^{\beta }$,$\ r=p$,$\ \alpha _{1}=\cdots =\alpha _{p}=0$, $\ m_{1}=\cdots
=m_{p}=m$ and $\lambda _{1}=\cdots =\lambda _{p}=-1$. Now, making appeal to
the integral representation (\cite{Mathai}, p.59): 
\begin{equation}
F_{A}^{r}\left( 
\begin{array}{c}
a,b_{1},\cdots ,b_{r} \\ 
c_{1},\cdots ,c_{r}%
\end{array}%
\mid x_{1},\cdots ,x_{r}\right) =\frac{1}{\Gamma (a)}\int_{0}^{+\infty
}e^{-t}t^{a-1}\ _{1}F_{1}\left(b_1; c_1;x_{1}t\right) \cdots \
_{1}F_{1}\left(b_r;c_r; x_{r}t\right) dt
\end{equation}%
where $\Re (a)>0$, together with the identity (\cite{Brychkov}, p.365): 
\begin{equation}
\frac{d}{dz}\ _{1}F_{1}(a;b;z)=\frac{a}{b}\ F\left(a+1; b+1; z\right)
\end{equation}%
of the confluent hypergeometric function $_{1}F_{1}$, one can use direct
calculations to obtain the derivative of $G_{p}$ as 
\begin{eqnarray}  \label{evaluating}
\frac{d}{dp}G_{p} &=&-\frac{ce^{\beta }}{p^{2}}\ F_{A}^{p}\left( 
\begin{array}{c}
1,-m,\cdots ,-m \\ 
1,\cdots ,1%
\end{array}%
\mid -\frac{\eta e^{\beta }}{p},\cdots ,-\frac{\eta e^{\beta }}{p}\right) 
\notag \\
&&-m\left( \frac{\eta e^{\beta }}{p}\right) ^{2}\ F_{A}^{p}\left( 
\begin{array}{c}
2,-m+1,-m,\cdots ,-m \\ 
2,1,\cdots ,1%
\end{array}%
\mid -\frac{\eta e^{\beta }}{p},\cdots ,-\frac{\eta e^{\beta }}{p}\right) .
\end{eqnarray}%
By evaluating $(\ref{evaluating})$ at $p=1$, we obtain 
\begin{equation}  \label{4.22}
\frac{d}{dp}G_{p}\big{|}_{p=1}=-\eta e^{\beta }{}_{1}F_{0}(-m;-\eta e^{\beta
})-m(\eta e^{\beta })^{2}{}_{1}F_{0}(-m+1;-\eta e^{\beta }).
\end{equation}%
Now, using the identity (\cite{Brychkov}, p. 381): 
\begin{equation}
_{1}F_{0}(a;z)=(1-z)^{-a},
\end{equation}%
Eq.$(\ref{4.22})$ reduces further to 
\begin{equation}
\frac{d}{dp}G_{p}\big{|}_{p=1}=-\eta \left( 1+\eta e^{\beta }\right)
^{m}-m(\eta e^{\beta })^{2}\left( 1+\eta e^{\beta }\right) ^{m-1}.
\end{equation}%
Therefore, the second integral in $(\ref{secondintegral})$ expresses as 
\begin{equation}
I_{2}=\frac{e^{-m\beta }}{-\eta e^{\beta }}\left( -\eta \left( 1+\eta
e^{\beta }\right) ^{m}-m\left( \eta e^{\beta }\right) ^{2}\left( 1+\eta
e^{\beta }\right) ^{m-1}\right) .  \label{I22}
\end{equation}%
By $(\ref{I11})$ and $(\ref{I22})$ we arrive at the announced formula $(\ref%
{entttropy})$. This ends the proof. $\Box $ \newline
\newline
\textbf{Proposition 4.3.} \emph{For $m\geq1, $ the minimum value of the
Wehrl entropy $(\ref{entttropy})$} is 
\begin{equation}
S_{W}\left(\hat{\rho}_{\beta_{\min}}\right)=1-\log\left(1-\tau_{m}\right)+m%
\left(-\log \ \tau_{m}+\tau_{m}-\tau^2_{m}\right)
\end{equation}
\emph{where } 
\begin{equation}
\tau_{m}:=\frac{1}{2}+\left(\frac{\sqrt{28+\frac{8}{m^3}+\frac{39}{m^2}-%
\frac{48}{m}}}{24\sqrt{3}}+\frac{m-1}{8m}\right)^{\frac{1}{3}}-\frac{2+m}{%
12m\left(\frac{\sqrt{28+\frac{8}{m^3}+\frac{39}{m^2}-\frac{48}{m}}}{24\sqrt{3%
}}+\frac{m-1}{8m}\right)^{\frac{1}{3}}},
\end{equation}
\emph{and} \emph{the temperature corresponding to this minimum is given by } 
\begin{equation}
T_{\min}=\frac{1}{\beta_{\min}}=\frac{-1}{\log(\tau_m)}.
\end{equation}
\ \newline
\textbf{Remark 4.2.} For $m=0$, the Wehrl entropy in $(\ref{entttropy})$
reduces to%
\begin{equation}  \label{swentropy}
S_{W}(\hat{\rho}_{\beta})=1-\log(1-e^{-\beta })
\end{equation}
which is a well-known result (\cite{Anderson}, p.2756) where the authors
pointed out that as the temperature $T\propto\beta^{-1}$ goes to zero $%
(\beta\to\infty)$, the uncertainty reduces to one which is the Lieb-Wehrl
bound expressing purely quantum fluctuations. This uncertainty becomes
larger tending to $\log T$ as the temperature goes to infinity. This limit
expresses purely thermal fluctuations. In \cite{Anderson}, comparison with
the von Neumann entropy for the same density operator, which reads (in
suitable unit) 
\begin{equation}
S_N(\beta)=-\log\left[2\sinh\left(\frac{\beta}{2}\right)\right]+\frac{\beta}{%
2}\coth\left(\frac{\beta}{2}\right).
\end{equation}
For large temperature, $S(\beta)\approx-\log(\beta)$, it coincides with $S_W$
in the high-temperature limit. But $S_N(\beta)\to 0$ as $T\to 0$ while $%
S_{W}(\hat{\rho}_{\beta})\to 1$ (see Figure 1). 
\begin{figure}[H]
\centering
\includegraphics[width=0.9\textwidth]{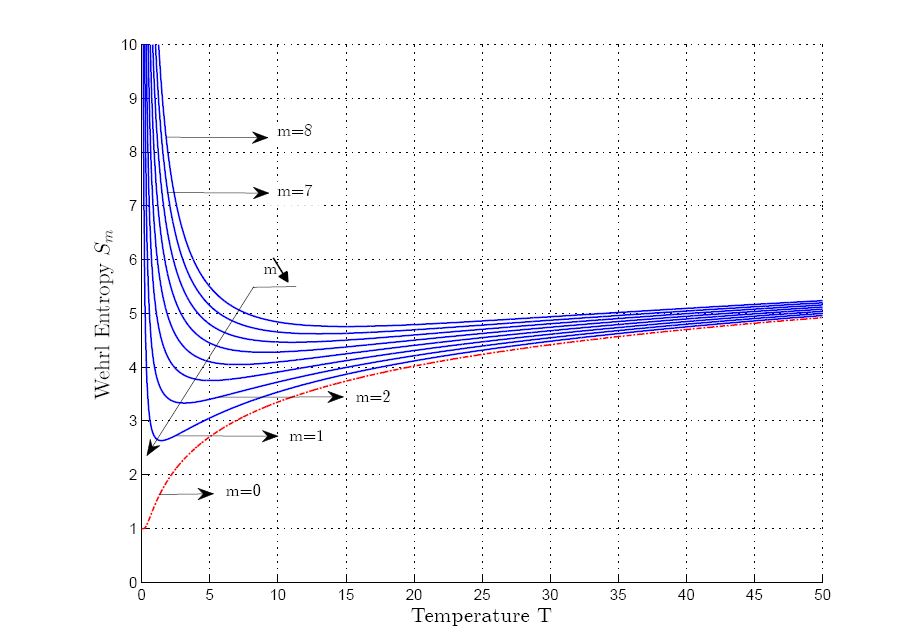}
\caption{Wehrl entropies $S_{W}\left(\hat{\protect\rho}_{\protect\beta%
}\right)$ as functions of $T=1/\protect\beta$ for $m=0,1,...8\ .$}
\label{}
\end{figure}
\begin{appendix}

\section{Proof of equation (3.2)} \label{appA}
We compute the scalar product
\begin{equation}
\langle \left( x,y\right) ,B,m\mid \varphi _{j}\rangle =\int_{\mathbb{R}} \langle \xi
\mid (x,y),B,m\rangle \overline{\langle \xi \mid \varphi _{j}\rangle }d\xi
\end{equation}
\begin{equation}\label{Husimi_Q}
=\left( \pi 2^{m+j}m!j!\right) ^{-1/2}e^{-\frac{B}{4}(x^{2}+y^{2})}\int_{%
\mathbb{R}}e^{-\left( \xi +\frac{\sqrt{B}}{2}(iy-x)\right) ^{2}}H_{m}(\xi
-\sqrt{B}x)\,H_{j}(\xi )\,d\xi
\end{equation}
By the change of variable $t=\xi +\frac{\sqrt{B}}{2}(iy-x)$, Eq\eqref{Husimi_Q} becomes
\begin{equation}
\langle (x,y),B,m\mid \varphi _{j}\rangle =\frac{e^{-\frac{B}{4}(x^{2}+y^{2})}}{\sqrt{\pi 2^{m+j}m!j!}}\int_{\mathbb{R}}e^{-t^{2}}\,H_{m}\left(t-%
\frac{\sqrt{B}}{2}(x+iy) \\
\right)\,H_{j}\left( t+\frac{\sqrt{B}}{2}(x-iy)\right)dt.
\end{equation}
The formula (\cite{Gradshteyn}, p.804):
\begin{equation}
\int_{-\infty }^{+\infty }e^{-t^{2}}H_{k}(t+u)H_{l}(t+v)\,dt=2^{l}\pi^{\frac{1}{2}}k!v^{l-k}L_{k}^{(l-k)}(-2uv),\,k\leq l,
\end{equation}
for parameters $u=-\frac{\sqrt{B}}{2}(x+iy)$ and $v=\frac{\sqrt{B}}{2}(x-iy)$ gives
\begin{equation}
\langle (x,y),B,m\mid \varphi _{j}\rangle =\sqrt{\frac{m!}{j!}}e^{-\frac{B}{4%
}(x^{2}+y^{2})}\left( \sqrt{\frac{B}{2}}(x-iy)\right)
^{j-m}L_{m}^{(j-m)}\left( \frac{B}{2}(x^{2}+y^{2})\right), \;m\leq j .
\end{equation}
In the same way, we obtain
\begin{equation}
\langle (x,y),B,m\mid \varphi _{j}\rangle =\sqrt{\frac{j!}{m!}}e^{-%
\frac{B}{4}(x^{2}+y^{2})}\left(-\sqrt{\frac{B}{2}}(x+iy)\right)
^{m-j}L_{j}^{(m-j)}\left( \frac{B}{2}(x^{2}+y^{2})\right),\;m\geq j .
\end{equation}
According to the expression of the Husimi function
\begin{equation}
Q^{(m)}_{j}((x,y),B)=\left| \left\langle (x,y),B,m|\varphi _{j}\right\rangle \right|
^{2},
\end{equation}
we get that
\begin{equation}
Q^{(m)}_{j}((x,y),B)=\frac{\min (m,j)!}{\max (m,j)!}e^{-\frac{B}{2}%
(x^{2}+y^{2})}\left( \frac{B}{2}(x^{2}+y^{2})\right) ^{|m-j|}\left( L_{\min
(m,j)}^{(|m-j|)}\left( \frac{B}{2}(x^{2}+y^{2})\right) \right) ^{2}.
\end{equation}
This ends the proof with $m\wedge j:=\min (m,j)$ and $m\vee j:=\max (m,j)$.$\Box$

\section{Proof of proposition 3.1.} \label{appB}
The characteristic function of the probability distribution $Q^{(m)}_{j}((x,y),B)$ is expressed as
\begin{equation}
\Phi_{Q^{(m)}_{j}}(u)=\frac{(m\wedge j)!}{(m\vee j)!}%
\int_{\mathbb{R}^2}\mathrm{e}^{(iu-1)\frac{B}{2}(x^{2}+y^{2})}\Big(\frac{B}{2}(x^{2}+y^{2})\Big)^{|m-j|}\Big(L_{m\wedge j}^{(|m-j|)}\Big(\frac{B}{2}(x^{2}+y^{2})\Big)^{2}\,d\mu(x,y)
\end{equation}
where $d\mu (x,y)=(2\pi)^{-1}Bdxdy$. Using the polar coordinates $x=r\cos
\theta, y=r\sin \theta,\ r>0, \theta
\in \lbrack 0,2\pi)$ and setting $\lambda =Br^{2}/2=B\left(x^{2}+y^{2}\right)/2$ we arrive at
\begin{equation}
\Phi_{Q^{(m)}_{_j}}(u)=\frac{(m\wedge j)!}{(m\vee j)!}%
\int_{0}^{+\infty} \mathrm{e}^{(iu-1)\lambda}\lambda^{|m-j|}\Big(L_{m\wedge j}^{(|m-j|)}(\lambda)\Big)^{2}\,d\lambda.
\end{equation}
Using the formula (\cite{Gradshteyn},p.809):
\begin{equation*}
\int_{0}^{+\infty }x^{\alpha }\mathrm{e}^{-bx}L_{m}^{(\alpha )}(\gamma
x)L_{n}^{(\alpha )}(\mu x)\,dx=\frac{\Gamma (n+m+\alpha +1)}{m!n!}\frac{%
(b-\gamma )^{m}(b-\mu )^{n}}{b^{n+m+\alpha +1}},
\end{equation*}
\begin{equation}
\times {}_{2}\!F_{1}\left( -m,-n;-m-n-\alpha ;\,b(b-\gamma -\mu )((b-\gamma
)(b-\mu ))^{-1}\right),
\end{equation}
where $\Re (\alpha )>-1$ and $\Re (b)>-1$, we obtain with $%
b=1-it,n=m\wedge j, \alpha=|m-j|$ and $\gamma=\mu=1$
\begin{equation}
\Phi _{Q^{(m)}_{j}}(u)=\frac{(m+j)!}{m!j!}\frac{(-1)^{m\wedge j}u^{2(m\wedge j)}%
}{(1-it)^{m+j+1}}\ {}_{2}F_{1}\left( -(m\wedge j),-(m\wedge
j);-m-j;(1+u^{2})u^{-2}\right).
\end{equation}
By the Pffaf's transformation (\cite{Magnus},p.47):
\begin{equation}
{}_{2}\!F_{1}(a,b;c;z)=(1-z)^{-a}\,{}_{2}\!F_{1}\left(
a,c-b;c;z(1-z)^{-1}\right) ,\quad z\in \mathbb{C},
\end{equation}
we obtain finally
\begin{equation}
\Phi _{Q^{(m)}_{_j}}(u)=\frac{(m+j)!}{m!j!}%
(1-iu)^{-(m+j+1)}\,{}_{2}\!F_{1}\left( -(m\wedge j),-(m\vee j);-m-j;\,1+u^{2}\right) .
\end{equation}
The moments of the variable $Q^{(m)}_{j}$ are obtained from the characteristic function repeating differentiation with respect to the variable $u$ and evaluated at the origin as
\begin{equation}\label{Exk}
\E((Q^{(m)}_{j})^{k})= \left. \frac{1}{i^k} \frac{\partial^k}{\partial u^k} \Phi_{Q^{(m)}_{j}}(u) \right|_{u=0}.
\end{equation}
The differentiation of the hypergeometric
function ${}_{2}\!F_{1}(\cdot )$ (\cite{Magnus},p.41) gives
the mean value
\begin{equation}
\mathbb{E}(Q^{m}_{j})=\frac{(m+j)!}{m!j!}(m+j+1)\,{}_{2}\!F_{1}\left(-(m\wedge j),-(m\vee j) ,-m-j;\,1+u^{2}\right)|_{u=0}.  \label{Exnp}
\end{equation}
We have from (\cite{Magnus},p.212):
\begin{equation}
{}_{2}\!F_{1}\left(-(m\wedge j),-(m\vee j) ,-m-j;\,1\right)=\frac{m!j!}{(m+j)!}.
\end{equation}
Then $ \mathbb{E}(Q^{(m)}_{j})=m+j+1$. For the variance, we need the second order moment of $Q^{(m)}_{\hat{\rho}_j}$ by applying \eqref{Exk} for $k=2$. We find
\begin{equation*}
\mathbb{E}\left((Q^{m}_{j})^{2}\right) =\frac{(m+j)!}{m!j!}(m+j+1)(m+j+2){}_{2}\!F_{1}\left(
-(m\wedge j),-(m\vee j),-m-j ;\,1+u^{2}\right) |_{u=0}
\end{equation*}
\begin{equation}
+\frac{(m+j)!}{m!j!}\frac{2mj}{m+j}\,{}_{2}\!F_{1}\left( -(m\wedge j-1),-(m\vee j-1),-(m+j-1);\,1+u^{2}\right) |_{u=0}  \label{Exn3}
\end{equation}
which reduces to
$
(m+j+1)(m+j+2)+2mj.
$
The well-known formula of the variance gives $ \mathcal{V}ar(Q^{(m)}_{j})=2mj+m+j+1.$ This ends the proof. $\Box$

\section{Proof of proposition 3.2.}\label{appC}
From \eqref{Qb_m}, the Husimi function of the mixed state $\hat{\rho}_{\beta}$ is expressed by
\begin{equation}
\label{hhusimi}
Q^{(m)}_{\beta}((x,y),B)=\frac{1}{Z}\sum_{j=0}^{+\infty }e^{-\beta
\left( j+\frac{1}{2}\right) }Q^{(m)}_{j}((x,y),B)
\end{equation}
According to remark 3.2, Eq.$(\ref{hhusimi})$ reads
\begin{equation}
\label{C2}
Q^{(m)}_{\beta}(\lambda,B)=\frac{e^{-\left(\lambda+\frac{\beta}{2}%
\right)}}{Z}\sum_{j=0}^{+\infty} e^{-\beta j} \frac{(m\land j)!}{(m\lor j)!}
\lambda^{|m-j|} \left(L_{m\land j}^{(|m-j|)}(\lambda) \right)^2.
\end{equation}
We start by computing the following series
\begin{equation}
\mathcal{O}_{\beta,\lambda,m} := \sum_{j=0}^{+\infty} e^{-\beta j} \frac{%
(m\land j)!}{(m\lor j)!} \lambda^{|m-j|} \left(L_{m\land
j}^{(|m-j|)}(\lambda) \right)^2
\end{equation}
\begin{equation*}
= \sum_{j=0}^{m-1} e^{-\beta j} \frac{j!}{m!} \lambda^{m-j}
\left(L_{j}^{(m-j)}(\lambda) \right)^2 - \sum_{j=0}^{m-1} e^{-\beta j} \frac{%
m!}{j!} \lambda^{j-m} \left(L_{m}^{(j-m)}(\lambda) \right)^2+ \sum_{j=0}^{+\infty} e^{-\beta j} \frac{m!}{j!} \lambda^{j-m}
\left(L_{m}^{(j-m)}(\lambda) \right)^2.
\end{equation*}
We denote by $S_{f}$ the finite sum
\begin{equation}
\label{C5}
S_{f} = \sum_{j=0}^{m-1} e^{-\beta j} \left[ \frac{j!}{m!} \lambda^{m-j}
\left(L_{j}^{(m-j)}(\lambda) \right)^2 - \frac{m!}{j!} \lambda^{j-m}.
\left(L_{m}^{(j-m)}(\lambda) \right)^2\right].
\end{equation}
Applying the formula (\cite{Brychkov2} p.552):
\begin{equation}
L_{m}^{(-k)}(x) = (-x)^k\frac{(m-k)!}{m!}L_{m-k}^{(k)}(x),\quad 1\leq k\leq
m,
\end{equation}
in the case $0\leq (k=m-j)\leq m$ since $j=0,1,\cdots m-1$, we find
\begin{equation}
L_{m}^{(j-m)}(\lambda) = (-1)^{m-j}\lambda^{(m-j)} \frac{j!}{m!}%
L_{j}^{(m-j)}(\lambda).
\end{equation}
Then $(\ref{C5})$ becomes
\begin{equation}
S_{f}= \sum_{j=0}^{m-1} e^{-\beta j} \left[ \frac{j!}{m!} \lambda^{m-j}
\left(L_{j}^{(m-j)}(\lambda) \right)^2 - \frac{j!}{m!} \lambda^{(m-j)}
\left(L_{j}^{m-j}(\lambda) \right)^2\right] = 0.
\end{equation}
So that $\mathcal{O}_{\beta,\lambda,m}$ takes the form
\begin{equation}
\mathcal{O}_{\beta,\lambda,m}=\frac{m!}{\lambda^m} \sum_{j=0}^{+\infty}
\frac{(e^{-\beta }\lambda)^j}{j!}\left(L_{m}^{(j-m)}(\lambda) \right)^2.
\end{equation}
With the help of Wicksell-Campbell-Meixner formula (\cite{Srivastava2}, p. 279):
\begin{equation}
\sum_{n=0}^{+\infty}\frac{u^n}{n!}L_l^{(n-l)}(\xi)L_k^{(n-k)}(\zeta)=e^u
\frac{u^l(u-\zeta)^{k-l}}{k!}L_l^{(k-l)}\left(-\frac{(\xi-u)(\zeta-u)}{z}%
\right)
\end{equation}
for parameters $l=k=m, n=j, \xi=\zeta=\lambda$ and $u=e^{-\beta}\lambda$, we obtain the expression
\begin{equation}
\mathcal{O}_{\beta,\lambda,m}=\exp\left(\lambda e^{-\beta}-\beta
m\right)L^{(0)}_m\left(-\lambda\left(1-e^{-\beta}\right)^2e^{\beta}\right).
\end{equation}
from which, \eqref{C2} becomes
\begin{equation}
Q^{(m)}_{\beta}(\lambda,B)=
\frac{1}{Z} \exp\left(\lambda(e^{-\beta}-1)-\beta\left(m+\frac{1}{2}%
\right)\right)L^{(0)}_m\left(-\lambda\left(1-e^{-\beta}\right)^2e^{\beta}\right).
\end{equation}
For the harmonic oscillator
$
Z=Tr\left(e^{-\beta\hat{H}}\right)=1/\left(e^{\beta/2}-e^{-\beta/2}\right)
$. This gives
\begin{equation}
Q^{(m)}_{\beta}(\lambda,B)=(e^{\beta/2}-e^{-\beta/2})\exp\left(%
\lambda(e^{-\beta}-1)-\beta\left(m+\frac{1}{2}\right)\right)L^{(0)}_m\left(-%
\lambda\left(1-e^{-\beta}\right)^2e^{\beta}\right).
\end{equation}
Finally, by replacing $\lambda$ by the expression $B(x^2+y^2)/2$, we obtain
the result. \ \ \ \ \ \ \ \ \ \ \ \ \ \ \ \ \ \ \ \ $\Box$

\section{Proof of proposition 4.1.}\label{appD}
The Wehrl entropy for the Husimi function \eqref{Husimi} is defined
by
\begin{equation}
\mathcal{S}_{W}(\hat{\rho}_j):=-\int_{\mathbb{R}^{2}}Q^{(m)}_{j}((x,y),B)\log Q^{(m)}_{j}((x,y),B)d\mu (x,y).
\label{WehrlJJ1}
\end{equation}
Using the polar coordinates $x=r\cos
\theta,\ y=r\sin \theta $ with $r>0$,  $\theta
\in \lbrack 0,2\pi)$ and setting $t =\frac{B}{2}r^{2}$, we rewrite $(\ref{WehrlJJ1})$ as
\begin{eqnarray*}
 \mathcal{S}_{W}(\hat{\rho}_j)&=& - \frac{(m\land j)!}{(m\lor j)!}\int_0^{+\infty}\left[\e^{-t}t^{|m-j|}\left(L_{m\land j}^{(|m-j|)}(t)\right)^2\right] \log \left[\frac{(m\land j)!}{(m\lor j)!} \e^{-t}t^{|m-j|}\left(L_{m\land j}^{(|m-j|)}(t) \right)^2 \right]dt
\end{eqnarray*}
We split this integral into four integrals as follows
\begin{eqnarray*}
\mathcal{S}_{W}(\hat{\rho}_j)&=& -\frac{(m\land j)!}{(m\lor j)!} \int_0^{+\infty}\left[\e^{-t}t^{|m-j|}\left(L_{m\land j}^{(|m-j|)}(t)\right)^2\right]dt \\ \nonumber
& & + \frac{(m\land j)!}{(m\lor j)!}\int_0^{+\infty}\left[\e^{-t}t^{|m-j|+1}\left(L_{m\land j}^{(|m-j|)}(t)\right)^2\right]\,dt  \\ \nonumber
& &- \frac{(m\land j)!}{(m\lor j)!}|m-j| \int_0^{+\infty}\left[\e^{-t}t^{|m-j|}\left(L_{m\land j}^{(|m-j|)}(t)\right)^2\right] \log t\,dt \\ & & - \frac{(m\land j)!}{(m\lor j)!}\int_0^{+\infty}\left[\e^{-t}t^{|m-j|}\left(L_{m\land j}^{(|m-j|)}(t)\right)^2\right] \log \left[\frac{(m\land j)!}{(m\lor j)!} \left(L_{m\land j}^{(|m-j|)}(t) \right)^2 \right]dt.
\end{eqnarray*}
The orthogonally relation of Laguerre polynomials gives
\begin{eqnarray}
I_1 &:=& - \frac{(m\land j)!}{(m\lor j)!}\int_0^{+\infty}\left[\e^{-t}t^{|m-j|}\left(L_{m\land j}^{(|m-j|)}(t)\right)^2\right]dt= -1 \label{I1}
\end{eqnarray}
From Srivastava's formula (\cite{Alassar},p.1133):
\begin{equation} \label{Srivastava}
\int_{0}^{+\infty} t^{\alpha+s} \e^{-t} \left(L_{n}^{(\alpha)}(t)\right)^2\, dt =  \sum_{r=0}^{n}\binom{s}{n-r}^2\frac{\Gamma(\alpha+s+r+1)}{r!}, \; \Re\alpha>-1.
\end{equation}
for $n=m\land j, \alpha=|m-j|$ and $s=1$, the seconde integral reads
\begin{eqnarray}
I_2 &:=& \frac{(m\land j)!}{(m\lor j)!}\int_0^{+\infty}\left[\e^{-t}t^{|m-j|+1}\left(L_{m\land j}^{(|m-j|)}(t)\right)^2\right]dt=m+j+1.
\end{eqnarray}
Next, the use of formula (\cite{Dehesa2},p.27):
\begin{eqnarray}
\int_{0}^{+\infty} \e^{-t} t^{\alpha+s} \left(L_{n}^{(\alpha)}(t)\right)^2 \log t \, dt &=&  \sum_{r=0}^{n}\binom{s}{n-r}^2\frac{\Gamma(\alpha+s+r+1)}{r!} \nonumber \\
&\times& \Big[2\psi(s+1)+\psi(\alpha+s+r+1)-2\psi(s+r+1-n)\Big]\nonumber
\end{eqnarray}
with $\Re\alpha>-1$ and $\psi(z)$ the digamma function, allows to write the third integral as follows
\begin{eqnarray}
I_3 :&=& -\frac{(m\land j)!}{(m\lor j)!}|m-j| \int_0^{+\infty}\left[\e^{-t}t^{|m-j|}\left(L_{m\land j}^{(|m-j|)}(t)\right)^2\right] \log t\,dt\\
 \label{I3}
&=& -|m-j|(m\lor j+1)\psi(m\lor j+1).
\end{eqnarray}
Making use of the formula (\cite{Dehesa3},p.3052):
\[
\frac{n!}{\Gamma(n+\alpha+1)} \int_{0}^{+\infty} \e^{-t} t^{s} \left(L_{n}^{(\alpha)}(t)\right)^2 \log \left[\frac{n!}{\Gamma(n+\alpha+1)} \left(L_{n}^{(\alpha)}(t)\right)^2\right] \, dt =  \frac{2^{2s+2}\Gamma(s+3/2)}{\sqrt{\pi}\Gamma(s+2)}n^{s+1} \]
\[
-\frac{2^{2s}(\alpha+1)\Gamma(s+1/2)}{\sqrt{\pi}\Gamma(s+1)}n^{s}\log n + \frac{2^{2s-1}\Gamma(s+1/2)}{\sqrt{\pi}\Gamma(s+1)}\Big[2(\alpha+1)\psi(s+1)-(2\alpha+1)\psi(s+1/2)-2\log \pi\]
\begin{equation}
-4(\alpha+1)\log 2+\gamma_E+4+2(\alpha+2s)+4\alpha s\Big] n^{s} + o(n^s)
\end{equation}
with $\Re \alpha>-1$ and $\gamma_E$ the Euler constant, the last integral expresses as
\begin{eqnarray}
I_4&:=&-\frac{(m\land j)!}{(m\lor j)!}\int_0^{+\infty}\left[\e^{-t}t^{|m-j|}\left(L_{m\land j}^{(|m-j|)}(t)\right)^2\right] \log \left[\frac{(m\land j)!}{(m\lor j)!} \left(L_{m\land j}^{(|m-j|)}(t) \right)^2 \right]\,dt \nonumber \\
& = & -m-j+(|m-j|+1)\log (m\land j)+\log2\pi-2 + o(1).
\end{eqnarray}
Finally, replacing the four integrals by their respective expressions, it follows that
\begin{eqnarray*}
\mathcal{S}_{W}(\hat{\rho}_{j})
&=& -|m-j|(m\lor j+1)\psi(m\lor j+1) + (|m-j|+1)\log (m\land j)+\log2\pi-1 + o(1).
\end{eqnarray*}
From  the asymptotic behavior of the digamma function ([31]):
\begin{equation}
\psi(m\lor j+1) = \log(m\lor j+1) + o(1)
\end{equation}
we obtain the announced result $(\ref{entrropy})$ and the inequality is immediate. This ends the proof.$\Box$
\end{appendix}


\end{document}